# Inorganic photovoltaic cells based on BiFeO$_3$: spontaneous polarization, lattice matching, light polarization and their relationship to photovoltaic performance

Chao He[1], Guocai Liu[1], Huiyan Zhao[2], Kun Zhao[1], Zuju Ma[3][**] and Xingtao An[1][*]

[1]*Department of Physics, School of Science, Hebei University of Science and Technology, Shijiazhuang, 050018, Hebei, China*

[2]*Department of Physics and Hebei Advanced Thin Film Laboratory, Hebei Normal University, Shijiazhuang, 050024, Hebei, China*

[3]*School of Materials Science and Engineering, Anhui University of Technology, Maanshan 243002, Anhui, China*

**ABSTRACT**

Inorganic ferroelectric perovskite oxides are more stable than hybrid perovskites. However, their solar energy harvest efficiency is not so good. Here, by constructing a series of BiFeO$_3$ based devices (solar cells), we investigated three factors that influence the photovoltaic performance, including spontaneous polarization, terminated ions species in the interface between BiFeO$_3$ and the electrode, and polarized light irradiation. This work was carried out in the framework of density functional theory combined with non-equilibrium Green's function theory under built-in electric field or finite bias. The results showed that 1) the photocurrent is larger only under a suitable electronic band gap rather than larger spontaneous polarization; 2) the photocurrent reaches the largest in Bi$^{3+}$ ions terminated interface than in the case of Fe$^{3+}$ or O$^{2-}$ with SrTiO$_3$ electrode; 3) the photocurrent could be largely enhanced if the polarized direction of the monochromatic light is perpendicular to the spontaneous polarization direction. The results would deepen the understanding of some experimental results of BiFeO$_3$ based solar cells.

---

[*][1] *Corresponding author. E-mail address*: anxingtao@semi.ac.cn
[**][2] *Corresponding author. E-mail address*: mazuju@fjirsm.ac.cn

## 1. INTRODUCTION

Since energy shortage is now a worldwide problem[1-4], researchers have begun to investigate the renewable energy materials, especially the solar energy harvest materials, such as solar cells. Efforts in the energy conversion efficiency improvement of silicon p-n junction solar cells are gradually loss of interest. The cost is relatively high and the efficiency could not exceed the Shockley-Queisser limit easily[5, 6], because the open circuit voltage could not exceed the value of band gap.

However, the recently widely studied perovskite materials do not encounter the open circuit voltage problem, theoretically. The open circuit voltage could exceed the band gap of the material and sometimes reached several dozens of volts[7]; thus, the energy conversion efficiency might exceed the Shockley-Queisser limit[8, 9]. Theoretically, the larger spontaneous polarization ($P_s$), the larger open circuit voltage in ferroelectrics based solar cell. This is because the electric field ($E$) is proportional to the spontaneous polarization value $P_s$ ($P_s = \chi\varepsilon_0 E$), and the open voltage ($U_o$) is proportional to $E$ ($U_o = cE$, here $c$ is the thickness of the device). We could draw the relationship, $U_o \propto E \propto P_s$, which was verified in experiments. For example, Bhatnagar et al.[10] found that the open circuit voltage of $BiFeO_3$ ($P_s$ = 80.2 μC/cm$^2$) could reach as large as 15 V, while Spanier's work[11] indicates that the open circuit voltage of $BaTiO_3$ ($P_s$ = 26 μC/cm$^2$) is about 4.8 V. The ratio of open circuit voltage is approximately equal to their ratio of spontaneous polarization, i.e. $U_o \propto P_s$.

According to depolarization theory[12-15], the above-mentioned electric field is mainly originated from the extra positive and negative charge in the two interfaces with the electrodes of the cell. This built-in electric field is also called depolarization field. The photovoltaic phenomenon in the ferroelectric perovskites is usually called bulk photovoltaic phenomenon (BPV)[16]. Excitons in ferroelectric perovskites are separated under this built-in electric field. While in the p-n junction based solar cells, the photo excited excitons are separated by the built-in electric field in the depletion layer. The depolarization field is inversely correlated with the thickness of the cell in experiment. The thicker of the cell in experiment, the more domains and defects, and the defects and domains would cancel out some intrinsic electric field in the perovskites[17, 18].

Some of these perovskites were the widely studied organic-inorganic hybrid halides perovskites. The energy conversion efficiency has reached to 22%. However, they have encountered stability problem and may pollute environment[19-21]. Inorganic ferroelectric perovskites rarely encountered stability problems. One of the candidate inorganic materials of the BPV effect perovskites is Bismuth Ferrite[1, 7, 10, 22-33] ($BiFeO_3$, or BFO). The extensively studied BFO is a multiferroic material which has the multiple properties of ferroelectric, anti-ferromagnetic, and ferroelastic[34]. The spontaneous polarization ($P_s$) in the hexagonal or rhombohedral BFO is about 80 μC/cm$^2$, which is nearly the largest room temperature value in the known perovskites[35].

However, the actual energy conversion efficiency of BFO is still too low (<2%)[3]. Subsequently, some experimental works are reported to enhance the photovoltaic performance of BFO based solar cells[8, 10, 15, 28, 29]. For example, Choi et al.[24, 30] have studied the relationship between electronic transport characteristics and the direction of spontaneous polarization by experimental methods. In addition, they have also studied the relationship between the photocurrent densities and polarized light. Interface effect is also a very key factor that influences the photovoltaic performance[22, 36]. However, systematically theoretical works that uncover these mechanisms of photovoltaic properties of BFO, especially the electrons transport properties based on the above three factors under sun illumination, are still rare.

In this work, we have done systematic first principles studies of BFO as a photovoltaic solar cell device to investigate the effect of $P_s$, terminated ions with electrode in the device, and directions of polarized light on the photovoltaic performance. With the help of density functional theory (DFT) combined with non-equilibrium Green's function (NEGF), the electronic density under bias or built-in electric field could be obtained. Subsequently, we could investigate how the photovoltaic performance changes by adjusting one main factor, whereas keeping other factors fixed as much as possible. The results may be helpful for the experiments to enhance the power conversion efficiency.

## 2. COMPUTATIONAL DETAILS

We firstly optimized the unit cell of BFO by Vienna Ab-initio Simulation Package (VASP) [37-40]. PAW method[37] was adopted to generate wave functions, and the interaction between ion cores and valence electrons was treated with ultra-soft pseudo potential[38]. The exchange and correlation effects of valence electrons were treated by generalized gradient approximation (GGA/PBE)[40] which is on the second order of Jacob's ladder[41]. The room temperature rhombohedral BFO is a G-type antiferromagnetic spin structure with Neel temperature 643 K[42]. Although the spin direction is a little tilted in adjacent, we only considered a collinear G-type antiferromagnetic model as other first-principles works to reduce the computational cost[42, 43]. So, we considered spin polarized calculation by SGGA. The strong correlation effect of Fe 3$d$ electrons is considered by the Coulomb repulsive energy $U_{eff}$ = 5 eV. Thus, we adopted SGGA + U[44, 45] method. This is the relatively computing resources inexpensive and suitable method, and it is widely adopted to deal with BFO rather than hybrid functional or GW.

Secondly, we constructed the devices based on the optimized structures by VASP. As shown in Fig. S1, the device is composed of left electrode, central region (including left electrode extension region, central scattering region, right electrode extension region), right electrode. Here, we assume that electronic transport direction is along $z$ axis. The device is two-dimensional periodic on the $xy$ plane and non-periodic in $z$-axis direction. The device calculations were performed with Atomistix ToolKit (ATK) package[46]. We adopted the linear combination of atomic orbitals (LCAO)[47] method to generate wave functions, and the interactions between ion cores and valence electrons was treated with OMX pseudo potential[48, 49], which is suitable to treat large systems with high accuracy. Here, we also adopted SGGA+U method to treat the exchange and correlation effect, spin polarization and strong correlation effect.

The easy polarization direction is along [1 1 1] for the room temperature BFO[35]. The $P_s$ is obtained with the theoretical framework of Berry phase[50], which is the modern theory of polarization. In this method, the charge center of dipoles in ferroelectrics are obtained by calculating the real space localized Wannier functions.

Under the built-in or applied electric field, the electron density in the device could not be obtained merely by DFT. But the electronic density could be obtained by DFT combined with NEGF method (DFT+NEGF)[46]. In this method, none equilibrium electron density is obtained by NEGF method, then the Hamiltonian is obtained by DFT method, the newly constructed Hamiltonian will then be put into the next NEGF calculations until the input and output charge densities are self-consistent.

The device Hamiltonian were divided into left part $H_L$, central part $H_{CR}$, and right part $H_R$,[46] which are respectively corresponding to left electrode, central region and right electrode in Fig. S1. $H_L$ and $H_R$ are calculated under periodic boundary conditions. The electron density of these two electrodes could be obtained by DFT. As for the central region, $H_{CR}$ is evaluated with open boundary conditions in the z-axis direction and periodic boundary conditions in the x-axis and y-axis directions. The corresponding spectral density matrix $\rho(E)^{L(R)}$ of the central region could be obtained by NEGF:

$$\rho(E)^{L(R)} = \frac{1}{2\pi} G(E) \Gamma(E)^{L(R)} G^\dagger(E) \qquad (1)$$

It contains the contributions from the left and right electrodes. $G(E)$ is the retarded Green's function matrix, and it contains the effect of the electrode states on the electronic structure of the central region, named self-energy[46] $\Sigma(E)^{L(R)}$:

$$G(E) = [(E + i\delta_+)S - H_{CR} - \Sigma(E)^L - \Sigma(E)^R]^{-1} \qquad (2)$$

Here, $\delta_+$ is an infinitesimal positive number, $S$ and $H_{CR}$ are the overlap and Hamiltonian matrices of the central region. The broadening function of the left (right) electrode is:

$$\Gamma(E)^{L(R)} = i[\Sigma(E)^{L(R)} - \Sigma(E)^{L(R)\dagger}] \qquad (3)$$

Once the spectral electron density is obtained, the new Hamiltonian $H_{CR}$ could be obtained by Kohn-Sham method. The final electron density could be obtained until the loop is self-consistent. Then, the related properties with the electron density in the device, such as, photocurrent, transmission spectrum and transmission eigenstates[51-53] will be obtained. This theoretical method has also been successfully applied to p-n junction solar cell[54].

For the central region with periodicity in x-axis and y-axis direction, both $H_{CR}$ and self-energy $\Sigma(E)^{L(R)}$ could be treated by Bloch expansion and corresponding k point sampling. The self-energy calculation needs a large number of k-points[55] in order to match the electronic structures of electrodes with the central region. In the entire device calculations above all, the k point's mesh is set to 2 × 2 × 100, and the energy convergence is set to be 1×10$^{-6}$ eV/atom.

In terms of the photocurrent calculations, the electron-photon interaction is given by the Hamiltonian[54, 56-58]:

$$H' = \frac{e}{m_0} \mathbf{A} \cdot \mathbf{P} \tag{4}$$

where $\mathbf{A}$ is the vector potential and $\mathbf{P}$ is the momentum operator. The electron-photon interaction $H'$ is included in $H_{CR}$. For a monochromatic light source, we have:

$$\mathbf{A} = \mathbf{e} \left( \frac{h\sqrt{\tilde{\mu}_r \tilde{\varepsilon}_r}}{4\pi N \omega c \tilde{\varepsilon}} F \right)^{1/2} \left( b e^{i\omega t} + b' e^{-i\omega t} \right) \tag{5}$$

Here, $\mathbf{A}$ is related to relative permeability $\tilde{\mu}_r$, relative permittivity $\tilde{\varepsilon}_r$, permittivity $\tilde{\varepsilon}$ of the material, and to the frequency $\omega$, photon flux $F$, number of photons $N$ and polarized direction of the light. In this paper, we adopted $\tilde{\mu}_r$= 3.2, obtained from experiments[59-61], and $\tilde{\varepsilon}_r$= 24 from our first principles calculations, the photon flux $F$ = 1s$^{-1}$Å$^{-2}$, and the polar light energy is ranged from 0 to 3.6 eV. So, we get the photocurrent of one electrode in the device (where $T^{\pm}_{\alpha,\beta}$ is the transmission coefficient of the two electrodes in opposite direction):

$$I_\alpha = \frac{e}{h} \int_{-\infty}^{+\infty} \sum_{\alpha,\beta=L,R} [1 - f_\alpha(E)] f_\beta(E - \hbar\omega) T^{-}_{\alpha,\beta}(E) - f_\alpha(E) [1 - f_\beta(E + \hbar\omega)] T^{+}_{\alpha,\beta}(E) dE \tag{6}$$

where $f_\alpha$ and $f_\beta$ are the corresponding Fermi distribution function. The electron-photon coupling matrix is contained in the above expression. Finally, the photocurrent formula is $I_{ph} = I_L - I_R$.

The transmission eigenstates are wave functions in the device under non-equilibrium condition, such as external or built-in electric fields. It is given by[55]:

$$\mathrm{T}(E) = Tr[\Gamma(E)^L G(E) \Gamma(E)^R G(E)^\dagger] \tag{7}$$

The contour value in the transmission eigenstates visualizing in this work is set to be 0.04 Å$^{-1.5}$eV$^{-0.5}$.

The electrostatic potential or Hartree potential $U_c$ (eV in unit) is obtained by solving the Poisson's equation of the electronic density in the crystal cells. Thus, we could calculate the built-in electric field (or depolarization field $E_d$) by the following formula:

$$E_d = \frac{\Delta U_c}{c} \tag{8}$$

where $c$ is the length of the central scattering region in the direction of the electric field. If the device was under finite bias, then effective potential (eV in unit) is needed to indicate if the electrons are easier or not to transmit in the device. It adds the potential under bias, exchange and correlation potential *etc*. to the electrostatic potential. Here the range of bias voltage is -1 ~ 1 volt.

The methods for calculating the photocurrent, effective potential and transmission eigenstates mentioned above are based on quantum mechanics, so the results would be more reliable.

## 3. RESULTS AND DISCUSSIONS

The room temperature BFO is not a standard cubic perovskite structure. It tilts from a structure with $Fm\bar{3}m$ cubic space group, and leads to $R3c$ hexagonal ($R3c$H) or rhombohedral ($R3c$R) crystal structure with Glazer notation[62] $a^-a^-a^-$ as shown in Fig. 1a. The driving force of structure distortion originates from the hybridization of O $2p$ orbitals and Fe $3d$ orbitals under the repulsive force of stereochemical active $Bi^{3+}$ 6s lone pairs[63]. We adopted the room temperature BFO as a prototype to construct the pure BFO solar cell devices, the Mn and Cr elements modulated $P_s$ solar cell devices (see Fig. 2a and 2b), and the $SrTiO_3$ (or STO) lattice matched solar cell devices (see Fig. 3a, 3b and 3c).

Due to the large computing resource consumption, the central scattering region of the device here is relatively smaller than the one in real device. We have made a convergence test about the effective potential of the BF(Cr,Mn)O and BF(Mn,Cr)O based devices (see Fig. S2). And we found that even if the central scattering region is smaller than the one in Fig. 2a, 2b, it does not change the relative potential barrier height.

To check the rationality of our theoretical methods adopted in this work, we computed the band gap and $P_s$ of bulk BFO, and the gap is 2.41 eV while the $P_s$ along [1 1 1] direction is 80.19 $\mu C/cm^2$. The results are consistent with previous theoretical work[43]. The calculated electrostatic potential difference in the $z$ direction of the perfect $R3c$H BFO cell is 0.976 V (This could be the open circuit voltage of the device, and consistent with the experimental results[64] at 0 K). Here, the lattice length in $z$ direction is $1.37 \times 10^{-9}$ m. According to Eq. (8), the calculated depolarization field is about $7.06 \times 10^8$ V/m, larger than experimental value $5 \times 10^7$ V/m[65], as is shown in Fig. 1d and lower panel of Fig. 1e. This is possibly because the device model that we adopted is perfect without considering complex domain structures in experiments[66-71]. The different directions of depolarization field in these domains could cancel out some components in $z$ direction overall. The depolarization field is the main origin of internal built-in electric field in this Nano device[14].

### 3.1 Polarization effects on the photocurrent

To investigate the effects of $P_s$ on the photovoltaic performance, let us consider such cells that keep the structures nearly unchanged, and have different $P_s$, in comparison with pure BFO. Fig. 1f and 1g give two such cells, Cr and Mn doped BFO cells (BF(Cr,Mn)O and BF(Mn,Cr)O). As shown in the upper panel of Fig. 1f, BF(Cr,Mn)O was constructed by using Cr and Mn to replace two adjacent Fe atoms in one BFO cell ($R3c$H) in $z$-axis direction, and BF(Mn,Cr)O in the upper panel of Fig. 1g was obtained by exchanging Mn and Cr positions of BF(Cr,Mn)O. The concentrations of Cr or Mn in the above two structures are all 3.3%. After optimization, the two cell structures are nearly unchanged. Fig. S3 and Tab. S1 show the bond length and bond angle of Cr (Mn) with the nearest O in the two cells. The valence states of Cr and Mn are all +3.

Here, Cr and Mn are chosen based on three factors as follows. Firstly, Cr and Mn have formed perovskite compounds with Bi, Fe and O in experiments, such as $Bi_2FeCrO_6$[72], $BiMnO_3$[73]. Secondly, Cr and Mn have similar ion radius with Fe. The cell structures are nearly unchanged after substituting Fe with Cr and Mn in BFO. Thirdly, Cr and Mn have different electro-negativity comparing with Fe. In the substituted BFO by Cr and Mn, there exists an additional electric field induced by Cr and Mn ions pairs, leading to the increase or decrease of the intrinsic electric field in pure BFO. As seen from the lower panel of Fig. 1f, the electric field of (Cr, Mn) pairs in BF(Cr,Mn)O is opposite with the intrinsic electric field in BFO, leading to the decrease of the intrinsic electric field in BFO. While the electric field of (Mn,Cr) pairs in BF(Mn,Cr)O is of the same direction with that of BFO, which enhance the intrinsic electric field in BFO (see the lower panel of Fig. 1g). The electric field in BF(Cr,Mn)O should thus be smaller than the one in BF(Mn,Cr)O. The different electric field corresponds to different $P_s$. The $P_s$ in BF(Cr,Mn)O is smaller than the one in BF(Mn,Cr)O.

Subsequently, two devices are constructed based on the above substituted hexagonal BFO structures. And the corresponding devices of the substituted cells of BFO are shown in the upper panel of Fig. 2b and 2a separately. The calculated $P_s$ is 78.47 μC/cm$^2$ for BF(Cr,Mn)O, and 87.12 μC/cm$^2$ for BF(Mn,Cr)O as shown in Tab. 1, verifying the above analysis for $P_s$ or the electric field in the BF(Cr,Mn)O and BF(Mn,Cr)O cells.

Larger depolarization electric field usually leads to larger electric current. Except for the depolarization electric field, the current is also related to other factors, such as the electrostatic potential barrier and carrier mobilities. As shown in Fig. 1d, the difference of electrostatic potential in BF(Cr,Mn)O is smaller than the one in BF(Mn,Cr)O. It means that the electrons in BF(Mn,Cr)O would move hardly than in BF(Cr,Mn)O. In addition, the calculated band gap (1.37 eV) for BF(Cr,Mn)O is smaller than the one (1.41 eV) for BF(Mn,Cr)O, as shown in Tab. 1. The smaller bandgap gives generally a larger electronic dielectric constant leading to a smaller exciton bonding energy because, as we can see in the Mott Wannier model[74, 75], the exciton bonding energy is directly related to the dielectric constant: $E_b=13.6\mu/\varepsilon^2$. It is indicated that the excitons in BF(Cr,Mn)O are easier to be separated by the depolarization electric field than in BF(Mn,Cr)O. That is to say that the carrier density in BF(Cr,Mn)O would be larger than the one in BF(Mn,Cr)O. So, the photovoltaic performance of BF(Cr,Mn)O based solar cell should be better than the one of BF(Mn,Cr)O based solar cell.

In order to illustrate this opinion, we calculated the photocurrent under AM1.5 (defined as the ratio between the length of sunlight propagation in the atmosphere $d$ and the atmosphere thickness $D$, here the ratio is 1.5) solar spectrum as shown in Fig. 2d. The photocurrent for BF(Cr,Mn)O based device presented in solid black line is larger than BF(Mn,Cr)O based one. The photocurrent density in BF(Cr,Mn)O based device under no bias voltage is $1.06\times10^2$ μA/cm$^2$, larger than $6.73\times10^1$ μA/cm$^2$ in BF(Mn,Cr)O based device as shown in Tab. 1. Although the device is under AM1.5 solar spectrum, only photons with energy larger than the band gap could be absorbed. This is because the electron hoping from valence band maximum (VBM) to conduction band minimum (CBM) needs to absorb photon energy larger than the band gap. The onset of the photocurrent is about 2.5 eV, larger than the corresponding band gap. Here, the band gap was calculated by the method of DFT, while the photocurrent was calculated mainly by NEGF. Both DFT and NEGF are the methods for solving the many-body Schrödinger equation. NEGF could give the results more precise than DFT[76]. The conventional exchange and correlation functionals such as GGA/PBE and LDA usually underestimate the band gap of some semiconductors due to the self-interaction error in the approximate exchange functional. So, the starting value of photocurrent is larger than the band gap, as shown in Fig. 2d. Here, we only care for the relative band gap, rather than the absolute one.

To uncover the reason why the $P_s$ value is not a key factor for the solar cell performance, we could also explain it in terms of the transmission eigenstates under finite bias (-1 ~ 1 volt) as shown in Fig. 2a and 2b on the lower panels. Obviously, the transmission eigenstates in BF(Cr,Mn)O based device spread nearly over the whole scattering region, see the lower panel in Fig. 2b, while they only gather in the extension region of the left electrode in BF(Mn,Cr)O based device, see the lower panel in Fig. 2a. From the perspective of effective potential, there exists a valley in the potential curve for the BF(Cr,Mn)O based device while there exists a peak for the one in BF(Mn,Cr)O based device as shown in Fig. 2c. The electrons could easily transport in the former one than the latter. So, the conductance in former device is better than the last one, because the excitons are easier to be formed and separated in the lower band gap materials. So, our result is similar to the case of organic-inorganic hybrid perovskites[77]. Our calculated photocurrent density is larger than experiments by Choi *et.al*[30]. This is because their experiments were adopted with only green light irradiation, while the photon energy is ranged from 0 to 3.6 eV in our whole theoretical work.

### 3.2 Interface atomic layer on the photocurrent density
This part mainly focuses on impact of special terminated ions (Bi, Fe and O) on the transport or photovoltaic performance of the device. So, we need to ignore some factors that do not influence the main conclusions, such as large lattice mismatch with metal, Coulomb screening at the interface of BFO and metal. We constructed a series of devices with different terminated ions for the right part of the device, where the interfaces are formed by (1 1 1) surface of BFO and (1 1 1) surface of STO, as shown in Fig. 3a, 3b and 3c. At the same time, we kept the left part of the device unchanged. Here, we chose STO instead of metal as electrode by considering the following factors. Firstly, STO is a perovskite oxide. The lattice mismatch between STO and BFO is smaller than other materials[78, 79], such as metal. If we use metal as electrode, many other factors that influent the photovoltaic performance would occur. Secondly, the band gap of STO is 3.25 eV[80], smaller than the widely known wide band gap semiconductor GaN (3.4 eV)[81]. The conductivity of STO should be higher than GaN. Thirdly, we chose STO as an electrode material based on some experimental works[82, 83].

We built the BFO/STO interface of the heterogeneous structure by Virtual NanoLab (VNL) tool used by Stokbro *et al*[84]. The interface was constructed when the average stress $\varepsilon^{av}$ is less than 0.65%. Here $\varepsilon^{av}$ is defined as the average stress of the three directions of the heterogeneous structure[85]. The stress in each direction is related to the rotation angle of the BFO and STO interface. At last, we choose the structure with the relatively lower average stress and strain energy. Thus, three devices with Bi, Fe, and O terminated ions were obtained. In addition, we kept the terminated ions of STO on the right part of the electrode unchanged in the above three cases.

In Bi terminated device (Device 1), the photocurrent density under no bias is larger than the one in O terminated device (Device 3), which is also larger than Fe terminated one (Device 2) as shown in Fig. 3e and Tab. 2. In order to find the reason, we firstly calculated the transmission eigenstates of the three devices as shown in Fig. 3a, 3b, and 3c. The transmission eigenstates in Device 1 spread almost over the whole device (see the lower panel of in Fig. 3a), while the states in Device 2 and 3 mainly focus on the edge of the BFO scattering region (see the lower panel in Fig. 3b and 3c). The results implicated that the conductance in the latter two devices are not so well compared to the first device. To uncover the origin of the phenomenon, we calculated the effective potential of electrons at finite bias, ranging from -1 to 1 volt as shown in Fig. 3d. Obviously, the potential barrier in Device 1 (blue dotted line) is lower than the one in Device 2, so the electrons could easily pass through it. For Device 2 (red dashed line) and Device 3 (green solid line), the potential barriers are all higher than the one of Device 1. Obviously, the electrons are easier to pass through Device 1 than Device 2 and Device 3. Due to the dramatic potential energy change near the right electrode in Device 2 (about 38 Å in *z* direction), the resistant force on the electrons (F $=-\partial V_{eff}/\partial z$) is larger than the one in Device 3. So, the conductance is not so well compared with Device 3, because the electrons are blocked on the left interface (see the lower panel in Fig. 3b). The large potential energy difference near the right electrode in Device 2 mainly originates from the big electronegativity difference between $Fe^{3+}$ in BFO and $O^{2-}$ in STO. At the same time, the photocurrent shows the similar tendency to the conductance of the devices. The different feature of these three devices is the different terminated ion species with the same STO electrode on the right part of these devices. Thus, the terminated ions of BFO have a great impact on the photovoltaic performance. We also deduced that the possible reason for the photovoltaic difference is the local dipole difference in the interfaces of the three devices. With the current level of experimental control, the detailed studies into the effect of the ions at the interfaces are not achievable.

### 3.3 Polarized light on the solar cell performance

For the anisotropic BFO ferroelectric materials, the polarized light should have an important impact on the excitons formation[29, 30]. In order to explain this phenomenon, two kinds of polarized light, which are $x$ direction polarized and $z$ direction polarized, were imposed on Device 2, and then corresponding transmission eigenstates and photocurrent were calculated as shown in Fig. 3b, 3e separately. The absolute value of photocurrent density was calculated and shown in Tab. 2. Obviously, the photocurrent density under no bias in the $x$ direction polarized light device is almost 10 times larger than the one in the $z$ direction polarized light device.

In order to explain this phenomenon, we calculated the projected band structure of bulk BFO along the high symmetry points in the Brillouin zone as shown in Fig. 1c and 1b separately. We found that the O 2$p$ orbitals dominate the VBM while Fe 3$d$ orbitals dominate the CBM and the effective mass of electrons (related to the curvature radius of the E - k curve) in valance band along $z$ direction (A→ Γ, M → L) is larger than the one in $x$ direction (Γ → K, H → A). The electrons under polarized light could easily hop from O 2$p$ orbitals to Fe 3$d$ orbitals in $x$ direction polarized case, because the electronic component in the electromagnetic wave is the main driving force to excite the electrons hoping from valence band to conduction band. Consequently, the photocurrent is larger in device under $x$ direction polarized light. Our theoretical work deepened the understanding of the experimental results by Choi *et al*[30].

## 4. CONCLUSIONS

In this article, we mainly investigated three key factors that influence the photovoltaic performance as solar cell by constructing a series of devices based on BFO. Firstly, although the photocurrent originates from the built-in electric field formed by electric dipoles, the larger $P_s$ is not the main factor that induces larger photocurrent, because the gap is another key factor for the carriers' formation. Smaller band gap gives generally a larger electronic dielectric constant, leading to a smaller exciton bonding energy, thus to larger photocurrent under the same solar illumination. Secondly, the interface ions of BFO have an important influence on the photocurrent when BFO is made into a heterogeneous solar cell, because the local electric field formed by interface ions could impact the electronic transport properties. This discovery inspired us that the photovoltaic performance could be enhanced not only by controlling the terminated ion species but also by contacting with electrode with different chemical potential. Thirdly, the energy conversion efficiency could be enhanced if the incident monochromatic light is polarized along a special direction. This phenomenon may originate from the anisotropy of carrier effective mass in the lattice of BFO. The results will deepen the understanding of some experimental results of BFO based solar cells.

### ACKNOWLEDGMENTS

This research was supported by National Science Foundation of China (No. 11747103, 21501177, 21771182, and 11575051), Hebei Funds for Distinguished

Young Scientists (No. A2018208076), Hebei Hundred Excellent Innovative Talents (No. SLRC2017035) and the Science Foundation of Hebei Education department for Young Scholar (No. QN2017086). We thank Dr. Yi Yang in Jiangxi University of Science and Technology for helpful discussions.

Tables:

**Tab. 1.** The electronic band gap, spontaneous polarization, total photocurrent, cross-sectional area, and photocurrent density of BF(Cr,Mn)O and BF(Mn,Cr)O based solar cell devices.

| Device | $E_g$(eV) | $P_s$(μC/cm$^2$) | $|I_{tot}|$(A) | $S$(cm$^2$) | $|j|$(μA/cm$^2$) |
|---|---|---|---|---|---|
| BF(Cr, Mn)O | 1.38 | 78.4 | 2.84×10$^{-19}$ | 2.69×10$^{-15}$ | 1.06×10$^2$ |
| BF(Mn, Cr)O | 1.41 | 87.1 | 1.81×10$^{-19}$ | 2.69×10$^{-15}$ | 6.63×10$^1$ |

**Tab. 2.** The total photocurrent, cross-sectional area, and current density of Device 1 (Bi terminated), Device 2 (Fe terminated) and Device 3 (O terminated) in the case of $z$ polarized polar light, and the corresponding parameters of Device 2 under $x$ direction polarized polar light.

| Device | $|I_{tot}|$(A) | $S$(cm$^2$) | $|j|$(μA/cm$^2$) |
|---|---|---|---|
| Device 1-$z$ | 7.80×10$^{-20}$ | 5.28×10$^{-15}$ | 1.48×10$^1$ |
| Device 2-$z$ | 2.93×10$^{-20}$ | 5.28×10$^{-15}$ | 5.55 |
| Device 2-$x$ | 3.26×10$^{-19}$ | 5.28×10$^{-15}$ | 6.17×10$^1$ |
| Device 3-$z$ | 6.16×10$^{-20}$ | 5.28×10$^{-15}$ | 1.17×10$^1$ |

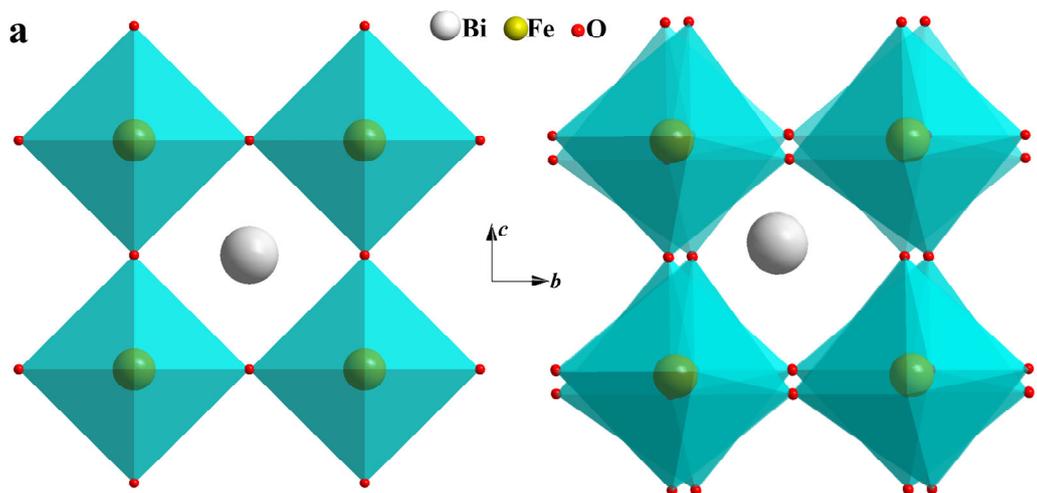
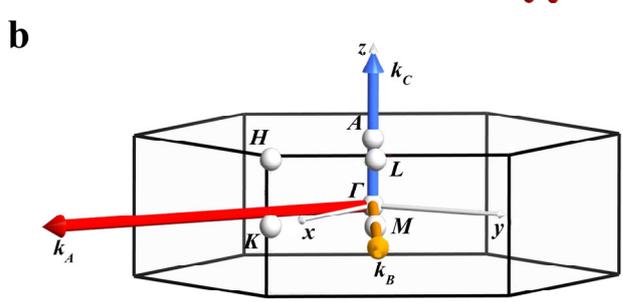
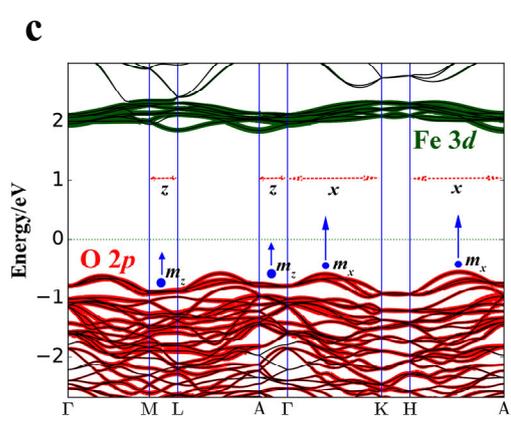
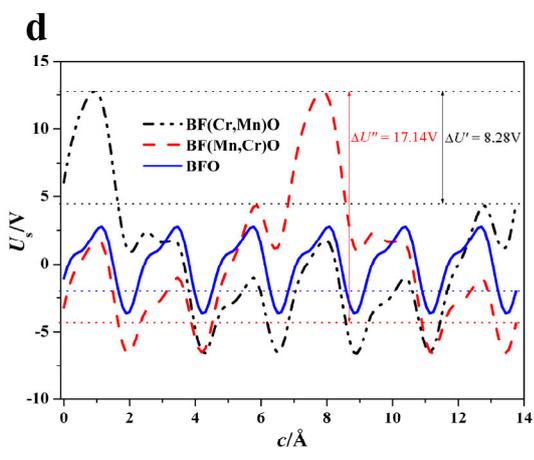
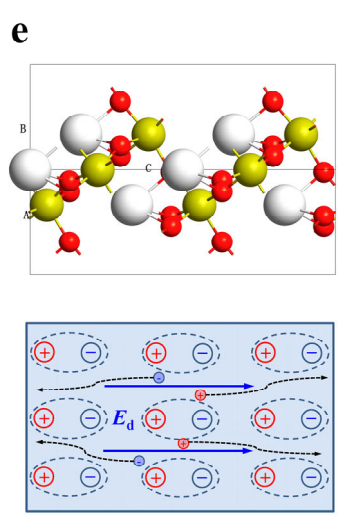
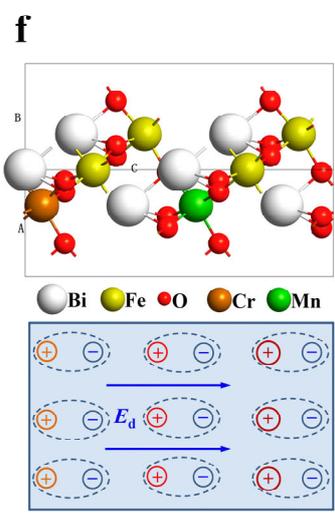
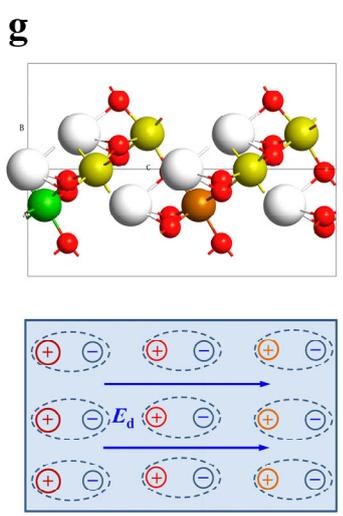

**Fig. 1**: Structure of perfect cubic BFO structure and rhombohedral distorted one with Glazer notation $a^-a^-a^-$ (a), Brillouin zone of $R3c$H lattice (b), the projected band structure (dark green lines mean Fe 3$d$ orbitals, red lines mean O 2$p$ orbitals) (c), the electrostatic potential of perfect BFO (solid blue line), BF(Mn,Cr)O (dashed red line), BF(Cr,Mn)O (dotted black line) based devices with no bias (d), unit cell of BFO (upper panel) and its depolarization model (lower panel) (e), unit cell of BF(Cr,Mn)O (upper panel) and its depolarization model (lower panel) (f), unit cell of BF(Mn,Cr)O (upper panel) and its depolarization model (lower panel) (g). In (e), (f) and (g), the length of the blue line shows the absolute value of depolarization field.

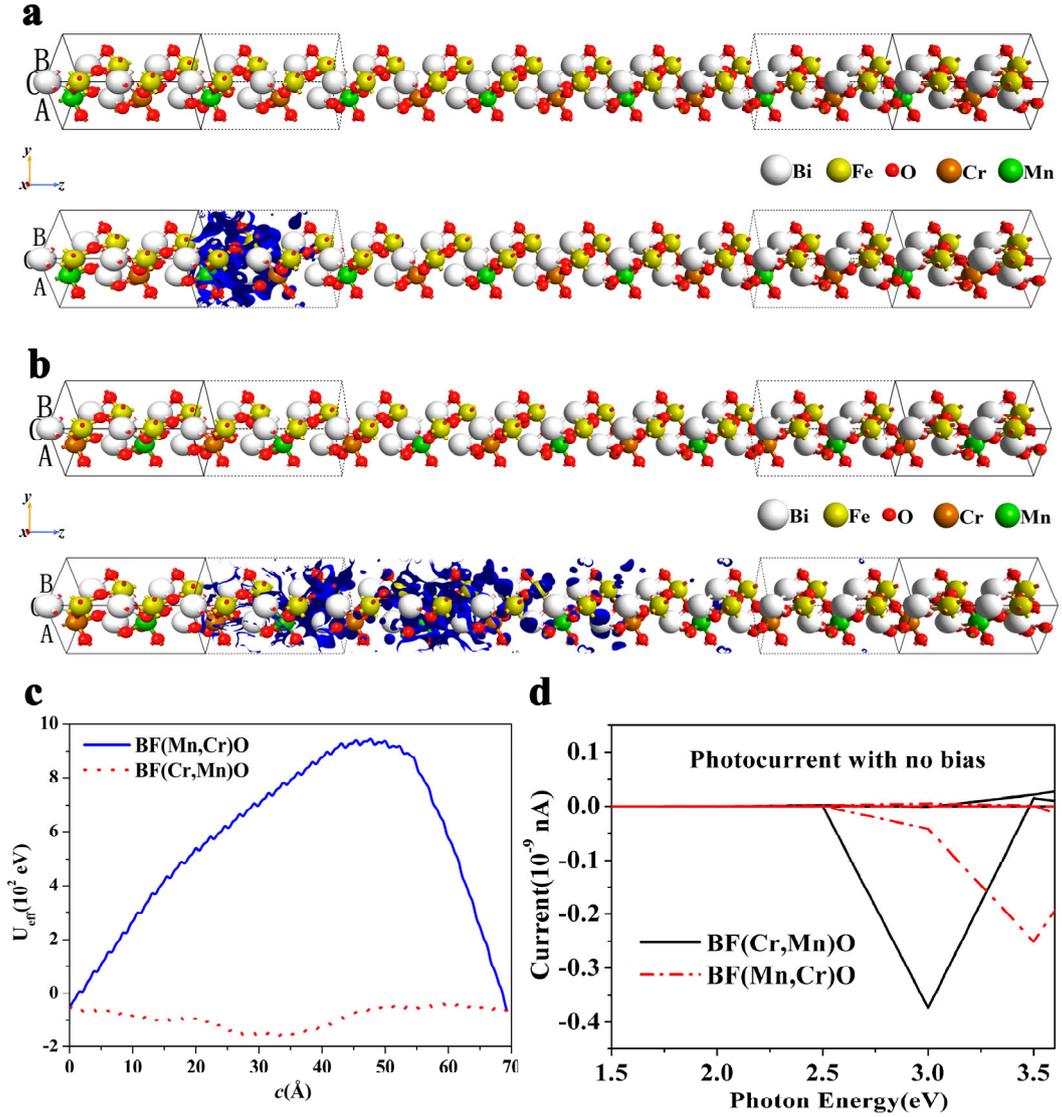

**Fig. 2**: The Mn, Cr substituted devices (BF(Mn,Cr)O) (upper panel) and related transmission eigenstates (lower panel) (a), and Cr, Mn substituted devices (BF(Cr,Mn)O) (upper panel) and related transmission eigenstates (lower panel) (b), the effective potential of BF(Mn,Cr)O with blue solid line and BF(Cr,Mn)O with dotted red line under finite bias (c), and related spin resolved photocurrent under AM1.5 solar spectrum with photon energy from 1.5 to 3.6 eV (d). The left and right leads of the two devices are infinite along *z* axis, and with a two-dimensional periodic boundary condition on the *xy* plane.

**Fig. 3**: Devices with Bi terminated (Device 1) and corresponding transmission eigenstates (a), Fe terminated (Device 2) and corresponding transmission eigenstates

(b), O terminated (Device 3) and corresponding transmission eigenstates (c), the corresponding effective potential under finite bias (d), and the spin resolved photocurrent with *z* polarized light and *x* polarized light (e).

Supporting Information

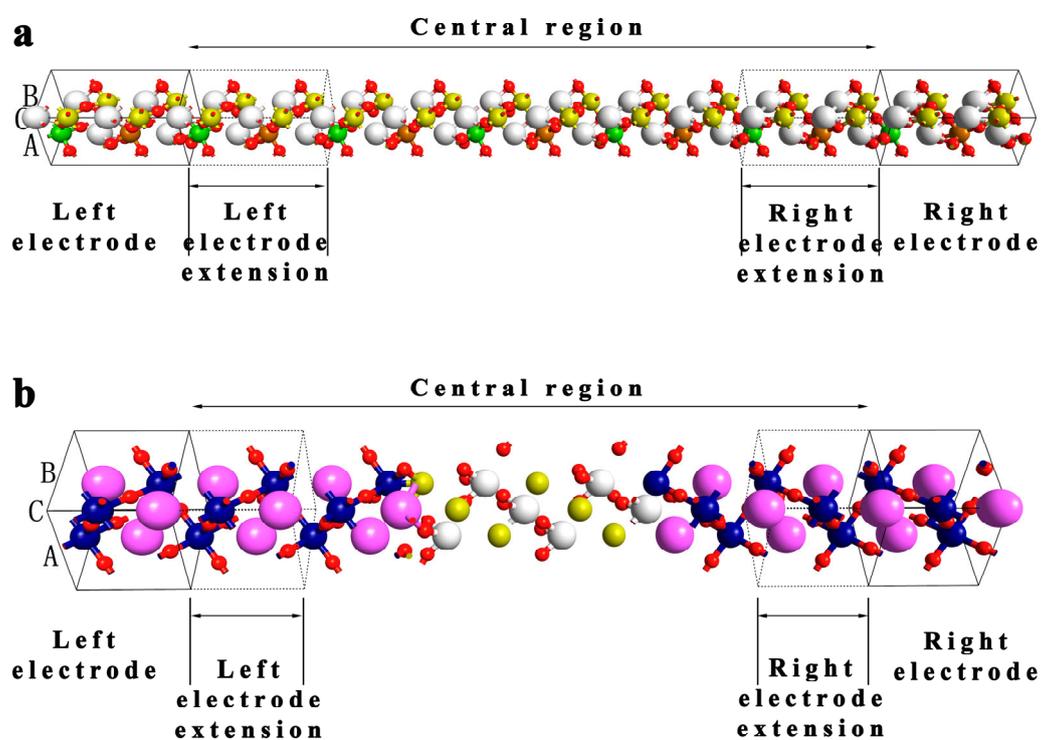

**Fig. S1**. The architecture of the device based on BF(Mn,Cr)O (a), and device composed of BFO and STO (b). They are all composed of left electrode, central region (including the left/right electrode extension, and central scattering region), and right electrode. For (b), the electrode is made of STO.

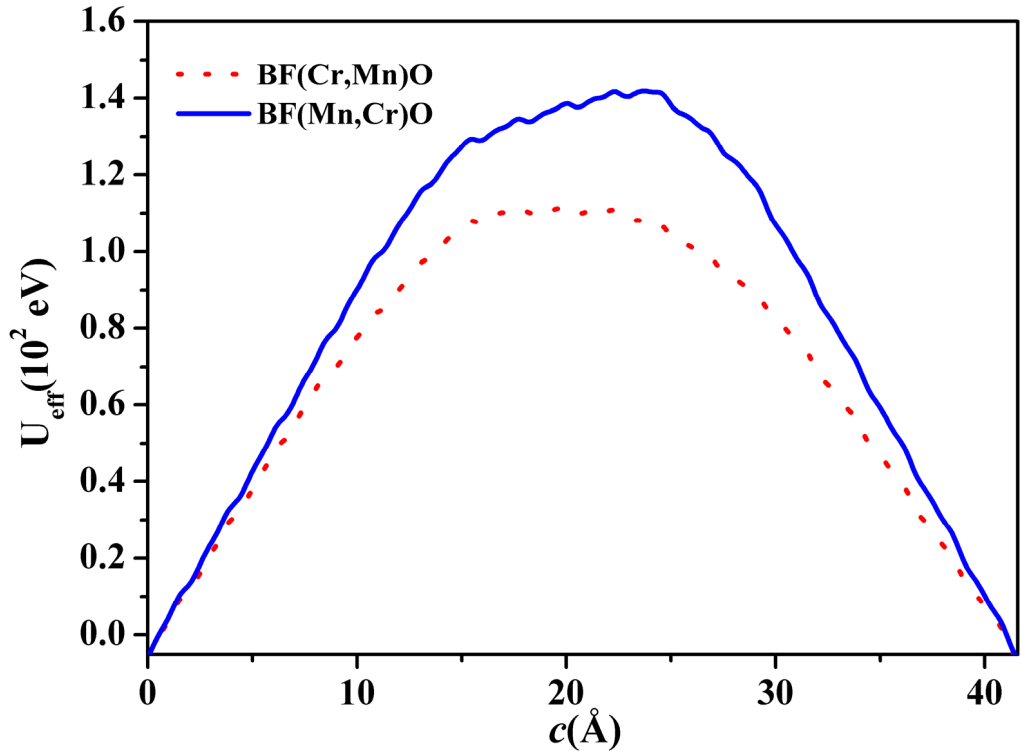

**Fig. S2** Effective potential under finite bias (-1~1 volt) for BF(Cr,Mn)O and BF(Mn,Cr)O based devices with the central region of 41.96 Å, smaller than the central region of the devices in Fig. 2 and Fig. 3 with the central region of 69.74 Å. The potential barrier of BF(Cr,Mn)O based device is still lower than BF(Mn,Cr)O based device, consistent with the conclusion in Fig. 2e.

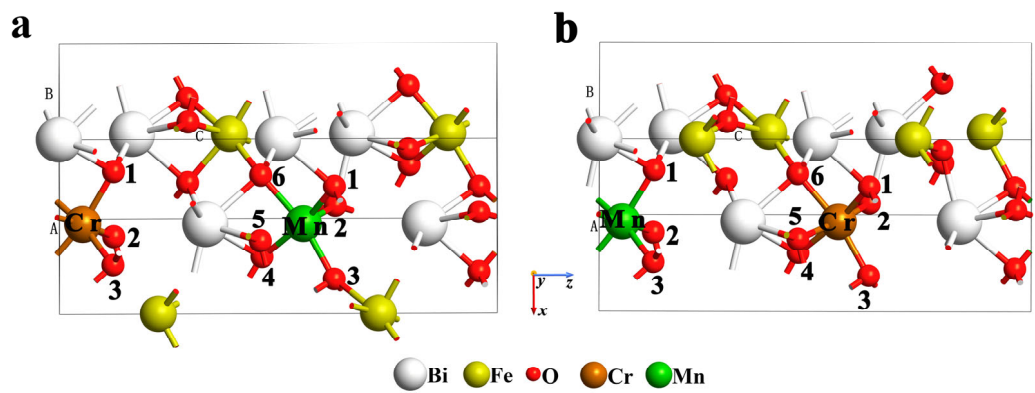

**Fig. S3**. The optimized cell structures of BF(Cr,Mn)O (a) and BF(Mn,Cr)O (b) by VASP. The element symbols of Cr and Mn are typed on the corresponding atoms. The nearest O atoms are distinguished by numbers.

**Tab. S1**. The bond length of Cr and Mn with the nearest O atoms and the corresponding bond angle in the cell of BF(Cr,Mn)O and BF(Mn,Cr)O.

|  |  | BF(Cr,Mn)O |  | BF(Mn,Cr)O |  |
|---|---|---|---|---|---|
| Bond length (Å) | Cr-O1 | 2.01 | Mn-O1 | 2.03 |
| | Cr-O2 | 2.01 | Mn-O2 | 2.01 |
| | Cr-O3 | 2.01 | Mn-O3 | 2.01 |
| | Mn-O1 | 2.05 | Cr-O1 | 2.02 |
| | Mn-O2 | 2.07 | Cr-O2 | 2.03 |
| | Mn-O3 | 2.05 | Cr-O3 | 2.02 |
| | Mn-O4 | 2.07 | Cr-O4 | 2.02 |
| | Mn-O5 | 2.10 | Cr-O5 | 2.05 |
| | Mn-O6 | 2.07 | Cr-O6 | 2.04 |
| Bond angle (°) | O1-Cr-O2 | 98.49 | O1-Mn-O2 | 97.70 |
| | O1-Cr-O3 | 98.90 | O1-Mn-O3 | 97.28 |
| | O2-Cr-O3 | 98.42 | O2-Mn-O3 | 97.63 |
| | O1-Mn-O2 | 98.53 | O1-Cr-O2 | 98.26 |
| | O1-Mn-O3 | 98.91 | O1-Cr-O3 | 98.56 |
| | O1-Mn-O5 | 89.65 | O1-Cr-O5 | 88.96 |
| | O1-Mn-O6 | 88.65 | O1-Cr-O6 | 88.44 |
| | O4-Mn-O2 | 88.21 | O4-Cr-O2 | 88.03 |
| | O4-Mn-O3 | 88.00 | O4-Cr-O3 | 88.41 |
| | O4-Mn-O5 | 82.64 | O4-Cr-O5 | 83.89 |
| | O4-Mn-O6 | 83.24 | O4-Cr-O6 | 83.77 |